\documentclass[aoas,MSNbibl,nameyear,noautosecdot,dvips]{arximspdf}
\usepackage{dcolumn}
\usepackage{graphicx}

\doi{10.1214/12-AOAS582} 
\volume{7}
\issue{1}
\pubyear{2013}
\firstpage{25}
\lastpage{50}

\makeatletter

\newcolumntype{d}[1]{D{.}{.}{#1}}

\newcommand{\rrvert}{\vert}
\newcommand{\llvert}{\vert}

\newcommand{\underset}[2]{\mathop{#2}_{#1}}

\makeatother

\begin{document}
\begin{frontmatter}

\title{Stronger instruments via integer programming in
an observational study of late preterm birth~outcomes}
\runtitle{Stronger instruments via integer programming}

\begin{aug}
\author[A]{\fnms{Jos\'{e} R.} \snm{Zubizarreta}\corref{}\ead[label=e1]{josezubi@wharton.upenn.edu}},
\author[A]{\fnms{Dylan S.} \snm{Small}\ead[label=e2]{dsmall@wharton.upenn.edu}},
\author[B]{\fnms{Neera K.} \snm{Goyal}\ead[label=e3]{neera.goyal@cchmc.org}},\\
\author[C]{\fnms{Scott} \snm{Lorch}\ead[label=e4]{lorch@email.chop.edu}}
\and
\author[A]{\fnms{Paul R.} \snm{Rosenbaum}\ead[label=e5]{rosenbaum@wharton.upenn.edu}}
\runauthor{J. R. Zubizarreta et al.}
\affiliation{University of Pennsylvania, University of Pennsylvania,
Cincinnati~Children's Hospital
Medical Center,\break The Children's Hospital of Philadelphia and
University of Pennsylvania}
\address[A]{J. R. Zubizarreta\\
D. S. Small\\
P. R. Rosenbaum\\
Department of Statistics\\
The Wharton School\\
University of Pennsylvania\\
Philadelphia, Pennsylvania 19104-6340\\
USA\\
\printead{e1}\\
\hphantom{E-mail: }\printead*{e2}\\
\hphantom{E-mail: }\printead*{e5}} 
\address[B]{N. K. Goyal\\
Department of Pediatrics\\
Cincinnati Children's Hospital Medical Center\\
Cincinnati, Ohio 45229\\
USA\\
\printead{e3}}
\address[C]{S. Lorch\\
Center for Outcomes Research\\
The Childrens Hospital of Philadelphia\\
Philadelphia, Pennsylvania 19104\\
USA\\
\printead{e4}}
\end{aug}

\received{\smonth{10} \syear{2011}}
\revised{\smonth{5} \syear{2012}}

%
\begin{abstract}
In an optimal nonbipartite match, a single population is
divided into matched pairs to minimize a total distance within matched
pairs. Nonbipartite matching has been used to strengthen instrumental
variables in observational studies of treatment effects, essentially by
forming pairs that are similar in terms of covariates but very
different in
the strength of encouragement to accept the treatment. Optimal
nonbipartite matching is typically done using network optimization
techniques that can be quick, running in polynomial time, but these
techniques limit the tools available for matching. Instead, we use integer
programming techniques, thereby obtaining a wealth of new tools not
previously available for nonbipartite matching, including fine and near-fine
balance for several nominal variables, forced near balance on means and
optimal subsetting. We illustrate the methods in our on-going study of
outcomes of late-preterm births in California, that is, births of 34 to 36
weeks of gestation. Would lengthening the time in the hospital for such
births reduce the frequency of rapid readmissions? A straightforward
comparison of babies who stay for a shorter or longer time would be severely
biased, because the principal reason for a long stay is some serious health
problem. We need an instrument, something inconsequential and haphazard
that encourages a shorter or a longer stay in the hospital. It turns out
that babies born at certain times of day tend to stay overnight once
with a
shorter length of stay, whereas babies born at other times of day tend to
stay overnight twice with a longer length of stay, and there is nothing
particularly special about a baby who is born at 11:00 pm. Therefore, we
use hour-of-birth as an instrument for a longer hospital stay. Using
integer programming, we form 80,600 pairs of two babies who are similar in
terms of observed covariates but very different in anticipated lengths of
stay based on their hours of birth. We ask whether encouragement to
stay an
extra day reduces readmissions within two days of discharge. A sensitivity
analysis addresses the possibility that the instrument is not valid as an
instrument, that is, not random but rather biased by an unmeasured covariate
associated with the hour of birth. Bias can give the impression of a
treatment effect when there is no effect, but it can also mask an actual
effect, leaving the impression of no effect, and both possibilities are
examined in analyses for effects and for near equivalence.
\end{abstract}

%

\end{frontmatter}

\section{\texorpdfstring{Introduction: Structure, application, data,
outline.}{Introduction: Structure, application, data, outline}}\label{secIntro}

\subsection{\texorpdfstring{The effects of changing the norms
for treatment.}{The effects of changing the norms for treatment}}

There are settings, common in medicine, clinical psychology and criminology,
in which certain norms govern the treatment assigned to an individual and
yet also a recognition that unique circumstances may justify a deviation
from the norm. In such a context, we might ask about the effects of
changing the norm without changing the latitude to deviate from the norm
when circumstances warrant a deviation. How should one study a situation
such as this?

In the current paper we look at late preterm births of 34 to 36 weeks
gestation in California and ask whether a shift in the norm for length
of stay in the hospital nursery reduces the frequency of rapid
readmission. Late preterm babies typically stay in the nursery for a
day or two before being discharged from the hospital. Should the norm
be one day or two days? Perhaps a two-day norm reduces the frequency of
rapid readmission, or perhaps one day is sufficient and the second day
is an unnecessary expense. Obviously, a baby with serious health
problems will and should be kept in the hospital as long as is
necessary---no one doubts the need to permit deviations from the
norm---and shifting the norm for a comparatively healthy baby is not
intended to alter the special care required by sick babies. We would
like to compare similar babies subject to different norms---one day or
two days---but with the same latitude to ignore the norm in specific
cases. A~straightforward comparison of babies who stay many days versus
babies who stay a single day will inevitably be a comparison of sick
and healthy babies and will provide no useful information about
changing the norm for healthy babies. \citet{GoyFagLor11} describe
changes over time in the norms for discharge of late preterm babies and
suggest that an evaluation of the effects of these changes is needed.

The question just raised---the question about changing the norm for
treatment while granting the same latitude for deviations from the
norm---is related to the so-called encouragement design [\citet
{Hol88}]; however, it asks a different question than is commonly asked
in that design. In a randomized encouragement experiment, some people
are picked at random and encouraged to take the treatment, while the
rest are not encouraged; however, there is noncompliance and people
often do not do what they are encouraged to do. Typically, in an
encouragement experiment, the goal is to estimate the effect of taking
the treatment, not the effect of being encouraged to take it, and
noncompliance is a nuisance whose consequences are to be removed
analytically. In the case of changing norms for treatment, deviations
from the norm are not properly called noncompliance, may be entirely
appropriate, even necessary, and we may have no interest in estimating
what would happen in a world which forbid deviations. No one wants to
discharge a sick baby who needs services provided by the hospital,
whatever norms are adopted for the length of stay of comparatively
healthy babies. How would outcomes change if the norms changed with no
change in the freedom to deviate from the norm? Notice that a change in
the norm might lead to a change in the way the freedom to deviate from
the norm is employed. Possibly, if the norm shifted from two days to
one day, more babies would deviate from the new one-day norm staying
instead the two days they would have stayed under the old two-day
norm.

In the case of norms, we are interested in the effects of changing the
encouragement without removing deviations from what is encouraged. In the
slightly specialized technical terminology introduced by \citet
{AngImbRub96}, we are interested in the causal effect of encouragement
on all
babies, not its effect on compliers, that is, the estimand of the numerator
of the Wald estimator, not the estimand of the Wald estimator itself.

\subsection{\texorpdfstring{Is a longer stay in the hospital nursery of
benefit to a newborn
baby\textup{?}}{Is a longer stay in the hospital nursery of benefit to
a newborn
baby}}\label{ssIntroExample}

The clock, the hour of birth, may alter whether a newborn baby stays in the
hospital nursery for one day or two before discharge to face the world for
the first time. In California, the typical baby born at 3:00 in the
afternoon (i.e., at 15:00) is discharged the following day, with a median
length of stay of 22 hours, while the typical baby born three hours
later at
6:00 in the evening (i.e., 18:00) is discharged after two days, with a median
length of stay of 43 hours. To the extent that the hour of birth is itself
inconsequential, to the extent that the hour of birth tells you nothing
about the health of the baby, it serves as an instrument, creating variation
in length of stay that will predict subsequent health outcomes only to the
extent that an extra day in the nursery is beneficial or harmful. See
\citet{AngImbRub96} for a general discussion of the use of
instrumental variables in causal inference.

An instrument is needed here because a straightforward comparison of babies
discharged earlier and those discharged much later is likely to be severely
biased. A~baby whose discharge is delayed for several days is very likely
to have significant complications requiring prolonged care or observation,
whereas a baby born at 6:00 in the evening is not an unusual baby.
Although biases are always conceivable in observational studies, there
is no
compelling reason to anticipate severe biases in a comparison of babies born
at 3:00 in the afternoon and others born at 6:00 in evening.

Briefly then, our plan is to form two subsets of babies using just the hour
of birth, those babies born at times that typically yield a one-day
stay and
those born at times that typically yield a two-day stay. More precisely,
we use hour of birth to produce pairs of babies with very different
anticipated lengths of stay (ALOS) based on hour of birth, specifically
based on the median length of stay for babies born at that hour. In other
words, we wish to focus attention on an innocuous source of variation in
length of stay, the hour of birth. Admittedly, our two groups do not
always stay one or two days, so our groups have heterogeneous lengths of
stay; however, unlike the hour of birth, variations in length of stay that
reflect the health of the baby are likely to bias comparisons of other
outcomes such as 2-day readmissions, and we do not want to use that portion
of the variation in length of stay in defining our comparison groups. See
\citet{MalBroKee00} and \citet{AlmDoy08} for related tactics.

An instrument is weak if it barely affects which treatment a baby receives
and it is strong if it is typically decisive in determining the treatment.
Weak instruments present substantial problems in part because they contain
little information [\citet{BouJaeBak95}] and in part because the
information
they do contain is sensitive to tiny unmeasured biases [\citet
{SmaRos08}]. Following the theory in \citet{SmaRos08} and extending
the technique in \citet{Baietal10}, we strengthen the instrument
by not
using all of the babies, forcing the remaining paired babies to be further
apart in terms of ALOS. Because the strength of an instrument affects its
design sensitivity, discarding some babies to increase strength can increase
the power of a sensitivity analysis [\citet{SmaRos08}]
despite the
contrary intuition that we all have from unbiased randomized experiments
where discarding observations can only reduce power.

The matching technique we use is a substantial advance over previous
techniques for this problem and more generally for so-called nonbipartite
matching problems. We use general integer programming techniques rather
than the subset of network optimization techniques. As reviewed in
Section~\ref{ssAlgorithmicBackground}, general integer programming
techniques are much
more flexible in what they can do, but in a certain abstract sense they are
not as suitable for large problems as are network optimization techniques.
Despite this abstract concern, we did not have difficulty in California
pairing 161,200 babies using integer programming, although the abstract
concern may be relevant in other practical contexts.

\subsection{\texorpdfstring{Data: Late preterms birth in California,
1993--2005.}{Data: Late preterms birth in California, 1993--2005}}

We used state\-wide discharge data on birth hospitalizations in California
from 1993 to 2005 obtained from the California Office of Statewide Health
Planning and Development. For each baby, there is a UB-92 form describing
principal diagnoses and medical procedures. These data were linked to
birth certificate data, maternal hospital records and hospital
admissions up
to one year after delivery. The data included live-born newborns delivered
vaginally at late preterm (34--36 weeks) gestation who were discharged
home. Using ICD-9-CM codes, we excluded newborns likely to require neonatal
intensive care because of major congenital anomalies, surgeries or
complications such as respiratory distress syndrome or sepsis. The clinical
team excluded newborns with length of stay $>$ 5 days, on the
grounds that prolonged hospitalization likely reflects significant
complications and possible neonatal intensive care.

\subsection{\texorpdfstring{Outline: A match, a matching algorithm,
an analysis.}{Outline: A match, a matching algorithm, an analysis}}

Section~\ref{secMatchedComparison} describes the matched comparison
while Section %
\ref{secInteger} discusses the optimization techniques used to create the
matched pairs. The optimization uses integer programming in a new way
on a
large scale. An analysis of one key outcome, readmission within two days
of discharge, is presented in Section~\ref{secInference}. The analysis tests
null hypotheses of both difference and near-equivalence and examines their
sensitivity to bias from unmeasured covariates [\citet{RosSil09N1}].
For instance, the analysis asks whether an apparent absence of effect
might be an effect of substantial magnitude masked by biases from unmeasured
covariates.

The manuscript presents an application, from conception through design
to analysis, but the novel methodological aspects are most prominent in
the construction of the matched pairs in Section~\ref{secInteger}.
These novel elements are easier to describe once the match has been
presented in Section~\ref{secMatchedComparison} and the distinction
between network and integer optimization has been reviewed in Section
\ref{ssAlgorithmicBackground}. The babies did not arrive as treated or
control babies; rather, the algorithm split one population of babies
into pairs so they have very different anticipated lengths of stay
based on the hour of birth; that is, in the technical terminology of
optimization theory, this is a nonbipartite match; for example, see
\citet{Edm65}, \citet{Der88} and Korte and Vygen
(\citeyear{KorVyg08}), Section 11. Nonbipartite matching has a variety
of uses in statistics [\citet{Luetal11}], for instance, matching
for time-dependent covariates [\citet{Lu05},
\citet{Siletal09}] and strengthening instrumental variables
[\citet{Baietal10}]. Concisely, if perhaps for the moment
obscurely, the novel elements of the integer programming algorithm in
Section~\ref{secInteger} include the following: (i) the extension of
fine balance to nonbipartite matching, including fine balance for
several variables at once, something that is not possible with network
optimization, (ii) the extension of optimal subset matching to
nonbipartite matching, (iii) the simultaneous use of fine balance and
optimal subset matching in nonbipartite matching, (iv) forcing balance
on means in nonbipartite matching. For a recent survey of the
literature on matching in observational studies, see
\citet{Stu10}.

\section{\texorpdfstring{The matched comparison: Similar covariates,
different anticipated
lengths of stay based on the hour
of birth.}{The matched comparison: Similar covariates, different anticipated
lengths of stay based on the hour of birth}}\label{secMatchedComparison}

For each hour of birth, 0 to 23, we computed the median length of stay in
the hospital. For instance, the median lengths of stay for babies born at
midnight, 11 am and 6 pm were, respectively, 37~hours, 26 hours and 43 hours.
Call this median length of stay for a given birth hour the ``anticipated
length of stay'' or ALOS. We formed 80,600 matched pairs of two similar
babies so that one baby in a pair had a much longer anticipated length of
stay than the other---at least 12 hours, and on average about 14
hours. Notice that these two groups of babies are defined by their individual
hours-of-birth, not their individual lengths of stay. We refer to these
paired babies as the ``long-hour-of-birth''
baby and the ``short-hour-of-birth'' baby
and abbreviate hour-of-birth as HOB. For instance, a baby born at 6 pm
might be paired with a baby born at 11 am, where the former would be the
long-HOB baby and the latter the short-HOB baby. The new algorithm we used
for this matching is described in detail in Section~\ref{secInteger},
but let us
first look at the resulting match, then consider its
construction.\looseness=-1

%
\begin{table}[b]
\tablewidth=240pt
\caption{Babies were matched exactly for 13 years of birth and for 311
hospitals, and this table displays the
counts of babies by year. Similar tables, not shown, for 311 hospitals
and $13 \times311$ years-by-hospitals
also exhibit perfect balance}\label{tabl1}
\begin{tabular*}{\tablewidth}{@{\extracolsep{\fill}}lcc@{}}
\hline
& \textbf{Long-HOB} & \textbf{Short-HOB}\\
\hline
\multicolumn{3}{@{}c@{}}{Year of birth, matched exactly} \\
[3pt]
1993 & 7471 & 7471 \\
1994 & 7514 & 7514 \\
1995 & 7221 & 7221 \\
1996 & 6877 & 6877 \\
1997 & 6644 & 6644 \\
1998 & 6191 & 6191 \\
1999 & 5814 & 5814 \\
2000 & 5702 & 5702 \\
2001 & 5505 & 5505 \\
2002 & 5348 & 5348 \\
2003 & 5547 & 5547 \\
2004 & 5416 & 5416 \\
2005 & 5350 & 5350 \\
\hline
\end{tabular*}
\end{table}

%
\begin{table}
\tablewidth=230pt
\caption{Balance for covariates that were either exactly matched or
finely balanced. The table counts
babies, and the total count in each subtable is 161\mbox{,}200 babies}\label{tabfine}
\begin{tabular*}{\tablewidth}{@{\extracolsep{4in minus 4in}}ld{6.0}d{6.0}@{}}
\hline
& \multicolumn{1}{c}{\textbf{Long-HOB}}
& \multicolumn{1}{c@{}}{\textbf{Short-HOB}} \\
\hline
\multicolumn{3}{@{}c@{}}{Birth $\mbox{weight}<2500$ grams, finely balanced} \\
[3pt]
$\ge$2500 grams & 72\mbox{,}500 & 72\mbox{,}500 \\
$<$2500 grams & 8100 & 8100 \\
[3pt]
\multicolumn{3}{@{}c@{}}{Gestational age, finely balanced} \\ [3pt]
34 weeks & 11\mbox{,}133 & 11\mbox{,}133 \\
35 weeks & 22\mbox{,}756 & 22\mbox{,}756 \\
36 weeks & 46\mbox{,}711 & 46\mbox{,}711 \\
[3pt]
\multicolumn{3}{@{}c@{}}{Gender, finely balanced} \\ [3pt]
Male & 42\mbox{,}549 & 42\mbox{,}549 \\
Female & 38\mbox{,}051 & 38\mbox{,}051 \\
[3pt]
\multicolumn{3}{@{}c@{}}{Race, finely balanced} \\ [3pt]
Hispanic & 40\mbox{,}342 & 40\mbox{,}342 \\
White & 24\mbox{,}067 & 24\mbox{,}067 \\
Asian & 7871 & 7871 \\
Black & 6009 & 6009 \\
Other & 2311 & 2311 \\
[3pt]
\multicolumn{3}{@{}c@{}}{Health insurance, finely balanced} \\ [3pt]
Federal & 42\mbox{,}061 & 42\mbox{,}061 \\
HMO & 31\mbox{,}461 & 31\mbox{,}461 \\
Fee for service & 3645 & 3645 \\
Uninsured & 2880 & 2880 \\
Other & 547 & 547 \\
Missing & 6 & 6 \\
[3pt]
\multicolumn{3}{@{}c@{}}{Parity, uniparous versus multiparous, finely
balanced} \\ [3pt]
Multiparous & 50\mbox{,}145 & 50\mbox{,}145 \\
Uniparous & 30\mbox{,}455 & 30\mbox{,}455 \\
[3pt]
\multicolumn{3}{@{}c@{}}{Multiple birth, finely balanced} \\
[3pt]
Single birth & 78\mbox{,}837 & 78\mbox{,}837 \\
Multiple birth & 1763 & 1763 \\
\hline
\end{tabular*}  \vspace*{-3pt}
\end{table}

%
\begin{table}
\tablewidth=270pt
\caption{Instrument imbalance and covariate balance in 80,600 matched
pairs of two babies, one
born at a long hour-of-birth (HOB), the other born at a short
hour-of-birth. The matching is intended
to construct pairs in which the anticipated length of stay (LOS) based
on the babies' hour of birth
is quite different, but covariates, such as birth weight, are similar.
Tabulated values are means.
Covariates are binary indicators except as noted}\label{tabcovs}
\begin{tabular*}{\tablewidth}{@{\extracolsep{\fill}}ld{4.2}d{4.2}@{}}
\hline
\textbf{Variable} & \multicolumn{1}{c}{\textbf{Long-HOB}}
& \multicolumn{1}{c@{}}{\textbf{Short-HOB}} \\
\hline
& \multicolumn{2}{c@{}}{Instrument} \\
[3pt]
Anticipated LOS (hours) & 39.56 & 25.48 \\
[3pt]
& \multicolumn{2}{c@{}}{Covariates} \\
[3pt]
Birth weight (grams) & 3064.96 & 3065.04 \\
High school degree & 0.60 & 0.60 \\
Birth injury & 0.01 & 0.01 \\
Oligohydramnios & 0.01 & 0.01 \\
Cord abnormality & 0.04 & 0.03 \\
Disorders of the placenta & 0.01 & 0.01 \\
Eclampsia & 0.00 & 0.00 \\
Chorioamniotis & 0.02 & 0.01 \\
Fever post-partum & 0.00 & 0.00 \\
Gestational diabetes & 0.03 & 0.03 \\
Diabetes mellitus & 0.00 & 0.00 \\
Prom & 0.09 & 0.07 \\
\hline
\end{tabular*}
\end{table}

The two babies in each pair were both born in the same year in the same
hospital, that is, the individual pairs were exactly matched for year
and hospital. Table~\ref{tabl1} shows the frequencies for the 13
years, and of course these are exactly the same for
the short-HOB and long-HOB babies. There is a similar exactly balanced
table, not shown, for the 311 hospitals, and a much larger exactly
balanced table, also not shown, for the interaction of year and
hospital with $13\times311=4043$ categories. Table~\ref{tabfine} shows
that the marginal distributions of seven other nominal variables were
exactly balanced, specifically birth $\mbox{weight} <2500$ grams,
gestational age, gender, race, health insurance, parity of the mother,
and single or multiple birth. (Because multiple births were very rare,
we make no special allowance for them.) Indeed, the exact balance seen
in Table~\ref{tabfine} is found within each hospital, that is, within
each of the $311$ categories. Unlike Table~\ref{tabl1}, Table
\ref{tabfine} exhibits fine balance, not exact pair matching; that is,
the marginal distributions seen in Table~\ref{tabfine} are exactly the
same, but within a single pair the two babies may differ
[\citet{RosRosSil07}]. However, we tried to pair individually
similar babies whenever possible [\citet{Zubetal11}]. Balance on
several other covariates is displayed in Table~\ref{tabcovs}.

Birth weight is the most important prognostic variable that is relevant to
all babies. For this reason, we matched for birth weight in four ways that
are described in detail in Section~\ref{secInteger}. Table
\ref{tabfine}
shows that the marginal distribution of low birth $\mbox{weight} <2500$ grams is
exactly balanced; this is a consequence of a fine balance constraint
[specifically (\ref{eqfinebalance}) in Section~\ref{secInteger}]. Also, Table
\ref{tabcovs}
shows the mean birth weights are reasonably close in the long and short HOB
groups (3064.96 grams for long-HOB and 3065.04 grams for short-HOB);
this is
a consequence of an approximate mean constraint [specifically (\ref
{eqmeanmatch1}) in Section~\ref{secInteger}]. The algorithm restricted the
number of babies mismatched for low birth weight [using (\ref{eqNearExact})
in Section~\ref{secInteger}] so that 97\% of pairs were individually
matched for
low birth weight; see Table
\ref{tablow}. Finally, an effort was made to pair individual babies
with similar birth
weights: the median absolute difference in weight for paired babies was 49
grams, and the upper quartile was 100 grams. The pairing of babies with
similar birth weights used a robust Mahalanobis distance that included birth
weight as one of the variables.

%
\begin{table}
\tablewidth=250pt
\caption{Matching for low birth $\mbox{weight} < 2500$ grams. The
marginal distributions are identical, as required by fine balance, and
$97\%$ of pairs are on the diagonal, exactly matched for birth
$\mbox{weight} < 2500$~grams. The table counts pairs, not
babies}\label{tablow}
\begin{tabular*}{\tablewidth}{@{\extracolsep{\fill}}ld{4.0}d{6.0}r@{}}
\hline
& \multicolumn{3}{c@{}}{\textbf{Short-HOB baby}} \\[-4pt]
& \multicolumn{3}{c@{}}{\hrulefill}\\
\textbf{Long-HOB baby} & \multicolumn{1}{c}{$\bolds{\le}$\textbf{2500 grams}}
& \multicolumn{1}{c}{$\bolds{\ge}$\textbf{2500 grams}}
& \multicolumn{1}{c@{}}{\textbf{Total}} \\
\hline
$\le$2500 grams & 6909 & 1191 & 8100 \\
$\ge$2500 grams & 1191 & 71\mbox{,}309 & 72\mbox{,}500 \\
[4pt]
Total & 8100 &72\mbox{,}500 & 80\mbox{,}600 \\
\hline
\end{tabular*}
\end{table}

We wanted the long-HOB baby and short-HOB baby to have very different
anticipated lengths of stay based on their hours of birth. The matching
algorithm began with all of the babies, splitting them into long and short
in an optimal manner while selecting an optimal subset to discard. Table
\ref{tabcovs}
shows that the average anticipated length of stay was 39.56 hours among
long-HOB babies and 25.48~hours among short-HOB babies.

%
\begin{table}[b]
\tablewidth=220pt
\caption{Actual days in the hospital in matched pairs. The table counts
pairs, not babies}\label{tabDays}
\begin{tabular*}{\tablewidth}{@{\extracolsep{\fill}}ld{6.0}rc@{}}
\hline
& \multicolumn{3}{c@{}}{\textbf{Short-HOB baby}} \\[-4pt]
& \multicolumn{3}{c@{}}{\hspace*{1pt}\hrulefill}\\
\textbf{Long-HOB baby} & \multicolumn{1}{c}{$\bolds{\le}$\textbf{1 day}}
& \multicolumn{1}{c}{\textbf{2 days}}
& \multicolumn{1}{c@{}}{$\bolds{\ge}$\textbf{3 days}} \\
\hline
$\le$1 day & 27\mbox{,}687 & 8704 & 1684 \\
2 days & 18\mbox{,}746 & 16\mbox{,}732 & 2061 \\
$\ge$3 days & 2443 & 1926 & \hphantom{0}477 \\
\hline
\end{tabular*}
\end{table}

\begin{figure}

\includegraphics{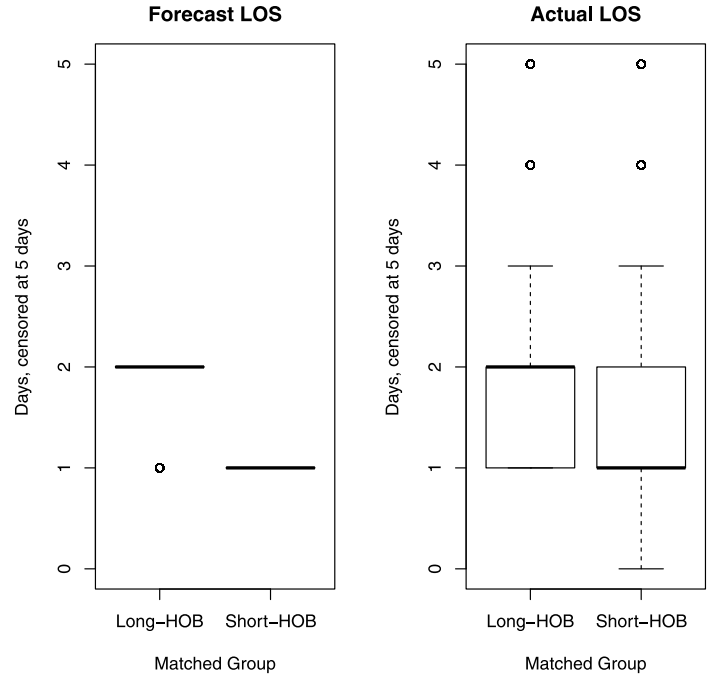}

\caption{Anticipated and actual length of stay (LOS) in days in 80,600
matched pairs of a long-HOB baby and a short-HOB baby. The anticipated
LOS for baby $ij$ is the median LOS for all babies with the same
hour of birth (HOB) as baby $ij$. The figure on the left shows
that babies in the long-HOB group were typically anticipated to stay two
days (36--60 hours) while babies in the short-HOB group were anticipated
to say one day (12--36 hours). The figure on the right is actual length
of stay.}\label{figur1}\vspace*{-3pt}
\end{figure}

How does anticipated length of stay based on hour of birth relate to actual
length of stay? Table
\ref{tabDays}
and Figure~\ref{figur1} provide answers. We defined zero days as less
%
%
than 12~hours, one day as between 12 and 36 hours, two days as between
36 and 50 hours, and so on, in effect rounding to the nearest 24 hour
unit. In Figure~\ref{figur1}, the boxplots on the left for anticipated
lengths of stay have collapsed into lines because the medians and
quartiles are equal: typically, long-HOB babies were anticipated to
stay two days and short-HOB babies were anticipated to stay one day. On
the right in Figure~\ref{figur1}, anticipation\vadjust{\goodbreak} often but not always
equaled actuality: the median and one quartile equaled the anticipated
stay. Presumably, the decision to keep a baby in the hospital for four
or more days in Figure~\ref{figur1} is not driven by the idiosyncrasy
of hour of birth, but rather by serious health problems of the newborn.
Table~\ref{tabDays} describes the actual length of stay in pairs.
Because babies were paired for important prognostic variables such as
birth weight, it is not surprising that the two babies in pair often
stayed the same number of days despite different hours-of-birth.
Nonetheless, in a pair, when one baby stayed two days and the other
stayed one, the odds were $18\mbox{,}746/8704=2.2$ to 1 that the long-HOB baby
was the one who stayed two days.

Section~\ref{secInteger} describes the new techniques used to construct this
match and Section~\ref{secInference} presents an illustrative analysis
of one
important outcome, namely, readmission to the hospital within two days of
discharge.

\section{\texorpdfstring{Using integer programming to construct the matched
comparison.}{Using integer programming to construct the matched
comparison}}\label{secInteger}

\subsection{\texorpdfstring{Some algorithmic background: Integer versus network
optimization.}{Some algorithmic background: Integer versus network
optimization}}\label{ssAlgorithmicBackground}

An integer programming problem is essentially a linear programming\vadjust{\goodbreak} problem
in which the solution is restricted to have integer coordinates rather than
fractional or real coordinates. Often, the solution is further restricted
to a subset of the integers, sometimes to 0 or 1. An excellent
introduction to integer programming is provided by \citet{Wol98}
and a more
detailed account is provided by \citet{Sch86}. Integer programs arise
in various problems in operations research because building $5.5$ submarines
and $6.5$ destroyers is actually less sensible than building 6 submarines
and 6 destroyers or 5 submarines and 7 destroyers or perhaps 8~submarines
and 5 destroyers. Integer programming shows up in optimal matching because
whole babies are matched to whole babies. Rounding the solution to a
linear program may be substantially inferior to solving an integer program,
but linear programming concepts play an important role in solving integer
programs.

An integer program has the form
%
%
\begin{equation}
\label{eqIntegerProgram}
\underset{\mathbf{a}}{\mathrm{minimize}}
\bolds{\eta}^{T}%
\mathbf{a}\quad\mbox{subject to}\quad\mathbf{Ba\leq b},
\mathbf{a}\geq\mathbf{0}\qquad\mbox{with }\mathbf{a}\mbox{ integer},
\end{equation}
where $\mathbf{B}$ is a given $d_{1}\times d_{2}$ matrix,
$\bolds{\eta}$ is a given $d_{2}$-dimensional vector with real
coordinates, $\mathbf{b}$
is a given $d_{1}$-dimensional vector, and one must find the best
$d_{2}$%
-dimensional vector $\mathbf{a}$ with $d_{2}$ integer coordinates. The
form (\ref{eqIntegerProgram}) simplifies the discussion in the current
section but, in general, integer programs may include both linear inequality
constraints (as in $\mathbf{Ba\leq b}$) and linear equality constraints (say,
$\mathbf{Ca=c}$) and, indeed, with a bit of juggling, either type of
constraint may be reexpressed in terms of the other, so a separate theory
for equality constraints is not needed. In the current paper and in most
matching problems $\mathbf{a}$ is further restricted to have binary, 1
or 0,
coordinates. The binary program is finite---there are $2^{d_{2}}$
candidate $\mathbf{a}$'s---but for large $d_{2}$ the number of candidates
suffers a combinatorial explosion and considering all of them, one by one,
is not possible. In the work here, $\bolds{\eta}$ and $\mathbf{a}$
have double subscripts, $\eta_{\ell m}$ and $a_{\ell m}$, $\ell<m$,
where $%
\eta_{\ell m}$ is a measure of distance on covariates between babies
$\ell$
and $m$, and $a_{\ell m}=1$ if babies $\ell$ and $m$ are paired and $%
a_{\ell m}=0$ if they are not. For instance, with $L$ babies, $\mathbf
{a}%
= ( a_{12},a_{13},a_{23},\ldots,a_{L-1,L} )^{T}$. Then $%
\bolds{\eta}^{T}\mathbf{a}$ is the total covariate distance within
matched pairs. The matrix $\mathbf{B}$ imposes various desired
restrictions on the match, not least that each baby shows up in at most one
pair.

In (\ref{eqIntegerProgram}), if you remove the restriction that
$\mathbf{a}$ is integral, then you have a linear program. The linear
program always has a minimizing value of
$\bolds{\eta}^{T}\mathbf{a}$ that is at least as small as the
integer program, but again that leaves you with the rather damp
prospect of half a submarine. There is a curious but important subset
of problems in which the linear programming solution and the integer
programming solution must be the same, and for these problems, known
somewhat inaccurately as network optimization problems, especially fast
algorithms are often available by adapting linear programming
techniques. These problems are called ``network optimization'' because
the most common versions arise from problems expressed in terms of the
nodes and arcs of graph theory. Somewhat more precisely, there is an
integral optimal solution to a linear programming problem if an integer
matrix $\mathbf{B}$ is totally unimodular, that is, if every square
submatrix of $\mathbf{B}$ has determinant $-1$, $0$, or $1$, a
condition that insures via Cramer's rule for matrix inversion
that linear equations solve with integer solutions. See Wolsey
[(\citeyear{Wol98}),
Section 3.2] for a precise statement and proof. In $\mathtt{R}$,
Hansen's (\citeyear{Han07})
\texttt{optmatch} package, Lu et~al.'s (\citeyear{Luetal11}) \texttt
{nbpmatching} package
and Yang et~al.'s (\citeyear{Yanetal}) \texttt{finebalance} package all use
network optimization
techniques, specifically the techniques of \citet{Ber81} and
\citet{Der88}. The restriction of $\mathbf{B}$ to be totally
unimodular is a
substantial restriction, and one can do quite a bit more with (\ref
{eqIntegerProgram}) if $\mathbf{B}$ is not so restricted, a fact we
demonstrate in detail in the current paper.

In abstract theory, solving large integer programs can be very
difficult. In particular, the general problem (\ref{eqIntegerProgram})
is NP-complete [\citet{Sch86}, Section~18.1]; however, specific
forms of (\ref{eqIntegerProgram}) are polynomially bounded [e.g.,
\citet{Sch86}, Section 18.6]. In practice, there has been a great
deal of progress in solving quite large integer programs either exactly
or approximately. We use IBMs ILOG program CPLEX to solve
(\ref{eqIntegerProgram}), and it is much faster than other programs we
have tried. IBM makes CPLEX available to academics for free.
\citet{Cor} created a package \texttt{Rcplex} that facilitates access
to CPLEX inside \texttt{R} and we have used \texttt{Rcplex} on Apple
and linux machines. In statistical matching, a common tactic is to
match exactly for a few key covariates [\citet{Ros10}, Section
9.3]---we did this for year and hospital---thereby breaking one large
matching problem into several smaller ones, each of which can be solved
quickly.

\subsection{\texorpdfstring{Nonbipartite matching using integer
programming.}{Nonbipartite matching using integer programming}}
\label{ssMatchingTechnique}

Generally, we wanted to match babies who were similar in terms of covariates
but very different in terms of anticipated length of stay based on hour of
birth. We matched exactly for hospital and year of birth, meaning that the
two babies in a pair were born in the same year at the same hospital.
Hospitals vary in discharge and readmission practices, so it was important
to compare two babies in the same hospital. There have been substantial
changes in discharge and readmission practices over the years, as well as
advances in medical technique, so matching for year was also important.
Exact matching can be implemented by simply dividing the population into
mutually exclusive and exhaustive subpopulations, and performing a separate
match for each subpopulation. Each subpopulaton consisted of a single
hospital over an interval of years. The rest of the discussion describes
the match within one such subpopulation, here a subpopulation defined by
hospital and year of birth.\looseness=-1

There are $L$ babies in the subpopulation, $\ell=1,\ldots,L$, and a
variable $a_{\ell m}$, $1\leq\ell<m\leq L$, with $a_{\ell m}=1$ if
babies $%
\ell$ and $m$ are paired and $a_{\ell m}=0$ otherwise. So $\mathbf{a}%
= ( a_{12},\ldots,a_{L-1,L} )^{T}$ has dimension
${L\choose2}$
and $\mathbf{B}$ has ${L\choose2}$ columns. The first constraint is
that $%
a_{\ell m}\in\{ 0,1 \} $ for all $\ell$, $m$, so the
problem is
not just an integer program but a binary program. Now each baby $\ell$
appears in at most one matched pair, and to enforce this, we impose $L$
linear inequalities, $\sum_{m=1}^{\ell-1}a_{m\ell}+\sum_{m=\ell
+1}^{L}a_{\ell m}\leq1$, for $\ell=1,\ldots,L$, which are coded as the
first $L$ rows of $\mathbf{B}$, where $b_{1}=1,\ldots,b_{L}=1$.

In statistics, matching\vspace*{1.5pt} is almost invariably ``without replacement,'' meaning
that no baby appears in more than one pair. The constraint $%
\sum_{m=1}^{\ell-1}a_{m\ell}+\sum_{m=\ell+1}^{L}a_{\ell m}\leq1$ ensures
matching is ``without replacement.'' Because outcome data are never
used in
constructing a match, when matching is without replacement, if the $L$
babies were independent prior to matching, then the pair outcomes are
conditionally independent in distinct pairs given the variables used to
construct the match, for instance, covariates and hour of birth. In
contrast, in matching ``with replacement,'' babies would be used
repeatedly in
different pairs, creating dependence. The analysis in Section~\ref{secInference}
uses existing techniques that are appropriate for conditionally independent
pairs, but these existing techniques are inapplicable when matching
``with
replacement.'' Indeed, even in the absence of bias from unmeasured
covariates, \citet{AbaImb08} argue that straightforward applications
of the bootstrap are inapplicable when matching ``with replacement,''
and that
the specialized techniques of \citet{AbaImb06} or \citet
{PolRom94} are required to obtain a standard error.

Suppose\vspace*{1.5pt} $L$ is even and one further equality constraint is added,
namely, $%
L/2=\sum_{m=1}^{L-1}\sum_{\ell=m+1}^{L}a_{m\ell}$ [so $\mathbf{B}$ has
an $%
L+1$ row consisting of a vector with ${L\choose2}$ coordinates all
equal to
1]. Then setting $\eta_{\ell m}$ equal to a covariate distance between
babies $\ell$ and $m$ and solving (\ref{eqIntegerProgram}) would yield a
minimum distance nonbipartite match that divides the $L$ babies into $L/2$
nonoverlapping pairs to minimize the total of the $L/2$ distances within
pairs. This optimization problem can be solved quickly using network
techniques [\citet{Der88}] as implemented in \texttt{R} in the
\texttt{%
nbpmatching} package [\citet{Luetal11}].

In contrast, the remainder of this section imposes additional
constraints as
additional rows of $\mathbf{B}$ to achieve specific effects, and these
require the integer programming formulation. In Section~\ref{sssFineBalance},
the marginal distributions of several nominal variables are forced to
balance exactly, a condition known as fine balance, as seen in Tables
\ref{tabl1} and
\ref{tabfine}. In Section~\ref{sssBinaryRequirements}, a binary
requirement is imposed on
pairs, while permitting a small fraction of pairs to escape the requirement
as needed, a condition which together with fine balance produced Table
\ref{tablow}
for low birth weight, with perfect balance for marginal distributions
combined with most pairs exactly matched. Section~\ref{sssBalancingMeans}
forces the means of a continuous covariate to balance, as seen for birth
weight in Table
\ref{tabcovs}, while Section~\ref{sssForcingMeanDifference} forces the
means of the instrument
to differ, thereby strengthening the instrument, as seen in Figure \ref
{figur1}. Fine
balance is generalized to near-fine\vadjust{\goodbreak} balance in Section \ref
{sssNearFine}. Finally, Section~\ref{sssOptimalSubset} adjusts $\eta
_{\ell m}$ to optimize
deletion of some babies while making the remaining babies closer on
covariates and further apart on the instrument.

In teaching, multiple linear regression is defined abstractly, and then
specific ways of coding its predictor matrix are shown to fit useful models,
such as polynomials or interactions. In parallel, the integer programming
solution to nonbipartite matching is best viewed abstractly as (\ref
{eqIntegerProgram}) with $a_{\ell m}\in\{ 0,1 \} $ and the
first $%
L$ rows of $\mathbf{B}$ requiring $\sum_{m=1}^{\ell-1}a_{m\ell
}+\sum_{m=\ell+1}^{L}a_{\ell m}\leq1$. Then one obtains a match that
meets specific requirements by suitably adjusting $\mathbf{B}$ and
$\eta_{\ell m}$, as described in Sections
\ref{sssFineBalance}--\ref{sssOptimalSubset}.

\subsubsection{\texorpdfstring{Fine balance.}{Fine balance}}\label
{sssFineBalance}

Table
\ref{tabfine}
exhibits fine balance of the marginal distributions for the seven nominal
variables. Fine balance for a covariate means that the marginal
distributions of the covariate are exactly the same in matched treated and
control groups, although individual pairs may not be exactly matched for
this covariate. If a nominal variable has $C$ categories, it is
represented as $C-1$ binary indicators. Let $w_{\ell}$ be the binary
indicator for one such category, say, $w_{\ell}=1$ if baby $\ell$ is
Hispanic and $w_{\ell}=0$ if baby $\ell$ is not Hispanic. Fine balance
for this category is the linear equality constraint
%
%
\begin{equation}
\label{eqfinebalance} \sum_{\ell=1}^{L-1}\sum
_{m=\ell+1}^{L}a_{\ell m} (
w_{\ell
}-w_{m} ) =0.
\end{equation}
Fine balance in Table
\ref{tabfine}
is actually present in every year in every hospital; that is, for instance,
among babies born in 2000 in hospital 22, the number of Hispanic long-HOB
babies equals the number of Hispanic short-HOB babies. Fine balance was
imposed through several linear equality constraints of this form. In
principle, an equality constraint (\ref{eqfinebalance}) may be
expressed\vspace*{1.5pt} in
the formulation (\ref{eqIntegerProgram}) as two inequalities or two
rows of $%
\mathbf{B}$, namely, $\sum_{\ell=1}^{L-1}\sum_{m=\ell+1}^{L}a_{\ell
m} ( w_{\ell}-w_{m} ) \leq0$ and $\sum_{\ell
=1}^{L-1}\sum_{m=\ell+1}^{L}a_{\ell m} ( w_{m}-w_{\ell} )
\leq
0 $; however, most solvers including CPLEX accept either inequality or
equality constraints. In CPLEX, each fine balance constraint (\ref
{eqfinebalance}) becomes one additional row of $\mathbf{B}$ with an equality
constraint.

Treated-versus-control minimum distance matching with fine balance for one
nominal variable, possibly with many levels, was proposed in Rosenbaum
[(\citeyear{Ros89}), Section 3.2] and \citet{RosRosSil07} using
either network
optimization or the optimal assignment algorithm; however, that
approach is
not applicable in nonbipartite matching and can only balance one nominal
variable. In contrast, the integer programming formulation of fine balance
(\ref{eqfinebalance}) is applicable to nonbipartite matching while balancing
one or more variables.

\subsubsection{\texorpdfstring{Binary requirements for individual
pairs.}{Binary requirements for individual pairs}}
\label{sssBinaryRequirements}

Let $h_{\ell m}\in\{ 0,1 \} $ be a binary variable
describing the
pairing of two babies, $\ell$ and $m$, where we wish to sharply limit the
number of times that paired babies\vadjust{\goodbreak} have $h_{\ell m}=1$, say, to at most $H$
pairs. Taking $H=0$ requires $h_{\ell m}=0$ for all $\ell$ and $m$,
whereas taking $H=5$ permits at most five matched pairs to have $h_{\ell
m}=1 $. In this study, we wanted paired babies to differ substantially in
terms of anticipated length of stay, so we set $h_{\ell m}=1$ whenever
baby $%
\ell$ had an anticipated length of stay that was less than 12 hours more
than the anticipated length of stay for baby $m$. The linear inequality
constraint
%
%
\begin{equation}
\label{eqNearExact} \sum_{\ell=1}^{L-1}\sum
_{m=\ell+1}^{L}a_{\ell m} h_{\ell m}
\leq H
\end{equation}
is added as a row to $\mathbf{B}$ to impose this constraint with $H=0$. In
addition, within each hospital in each year, a constraint of the form
(\ref{eqNearExact}) was used with $h_{\ell m}=1$ if babies $\ell$ and $m$
differed in terms of low birth weight $<2500$ grams and $H$ was twenty
percent of the number of births in that hospital in that year.

\subsubsection{\texorpdfstring{Balancing means.}{Balancing means}}\label
{sssBalancingMeans}

For any covariate $v$, not necessarily a binary covariate, suppose that we
wish to ensure that the means in matched treated and controls groups differ
by at most a number $\varepsilon>0$. Unlike a binary covariate in (\ref
{eqfinebalance}), for a continuous covariate such as birth weight, one
cannot reasonably take $\varepsilon=0$. Because there are $\sum_{\ell
=1}^{L-1}\sum_{m=\ell+1}^{L}a_{\ell m}$ matched pairs, this
requirement is
the same as
%
%
\begin{equation}
\label{eqmeanmatch1} \Biggl\llvert\sum_{\ell=1}^{L-1}
\sum_{m=\ell+1}^{L}a_{\ell m}
v_{\ell
}-\sum_{\ell=1}^{L-1}\sum
_{m=\ell+1}^{L}a_{\ell m} v_{m}\Biggr
\rrvert\leq\varepsilon\sum_{\ell=1}^{L-1}\sum
_{m=\ell+1}^{L}a_{\ell m}.
\end{equation}
Now, because of the absolute values in the constraint (\ref{eqmeanmatch1}),
this constraint is not one linear inequality. However, requiring (\ref
{eqmeanmatch1}) to hold is equivalent to requiring two linear inequalities
to both hold, namely,%
%
%
\begin{equation}
\label{eqmeanmatch2}\sum_{\ell=1}^{L-1}\sum
_{m=\ell+1}^{L}a_{\ell m} (
v_{\ell
}-v_{m}-\varepsilon) \leq0\quad\mbox{and}\quad\sum
_{\ell
=1}^{L-1}\sum_{m=\ell+1}^{L}a_{\ell m}
( v_{m}-v_{\ell
}-\varepsilon) \leq0.\hspace*{-24pt}
\end{equation}
So, a requirement that the means of $v$ after matching differ by at
most $%
\varepsilon$ is represented in the integer program as two rows of the
matrix $\mathbf{B}$. Notice in Table
\ref{tabcovs}
that the mean of birth weight is almost the same for the long-HOB and
short-HOB babies. The same technique was applied to birth injury and
oligohydramnios in Table~\ref{tabcovs}.

\subsubsection{\texorpdfstring{Near-fine balance.}{Near-fine
balance}}\label{sssNearFine}

Sometimes fine balance (\ref{eqfinebalance}) for a binary variable $w$
is infeasible or just too restrictive. For bipartite matching,
\citet{Yanetal} proposed a network optimization algorithm for
treatment-versus-control near-fine balance requiring $\llvert\sum_{\ell
=1}^{L-1}\sum_{m=\ell+1}^{L}a_{\ell m} ( w_{\ell}-w_{m} )
\rrvert\leq\varepsilon$ rather than (\ref{eqfinebalance}) for the
binary variables $w$ that define categories of a single nominal
variable, and Yang implemented this in her \texttt{finebalance} package
in \texttt{R} which uses network optimization. Just as
(\ref{eqmeanmatch1}) became two linear inequalities in
(\ref{eqmeanmatch2}), so too $\llvert\sum_{\ell
=1}^{L-1}\sum_{m=\ell+1}^{L}a_{\ell m} ( w_{\ell}-w_{m} )
\rrvert\leq\varepsilon$ may be split into two linear inequality
constraints which are imposed using integer programming. Also, unlike
network optimization, integer programming permits near-fine balance for
one or more nominal variables in nonbipartite matching.

\subsubsection{\texorpdfstring{Forcing pairs to differ with respect to
the mean of the
instrument.}{Forcing pairs to differ with respect to the mean of the
instrument}}\label{sssForcingMeanDifference}

Although we set a minimum requirement of a 12 hour difference in anticipated
length of stay using a constraint of the form (\ref{eqNearExact}), we wanted
the typical difference to be larger than the minimum. Specifically, we
imposed the requirement that the mean difference in anticipated length of
stay, say, $v_{\ell}$, should be at least $\phi=13$ hours, that is, we
required
\[
\sum_{\ell=1}^{L-1}\sum
_{m=\ell+1}^{L}a_{\ell m} v_{\ell}-\sum
_{\ell
=1}^{L-1}\sum_{m=\ell+1}^{L}a_{\ell m}
v_{m}\geq\phi\sum_{\ell
=1}^{L-1}\sum
_{m=\ell+1}^{L}a_{\ell m}
\]
by imposing the linear inequality constraint%
\[
\sum_{\ell=1}^{L-1}\sum
_{m=\ell+1}^{L}a_{\ell m} ( v_{\ell
}-v_{m}-
\phi) \geq0.
\]
In Table
\ref{tabcovs}, the anticipated length of stay based on birth hour is
39.56 hours for the
long-HOB babies and 25.48 hours for the short-HOB babies, an anticipated
difference of more than 14 hours.

\subsubsection{\texorpdfstring{Using several techniques to balance one
covariate.}{Using several techniques to balance one covariate}}

It is possible to use several of these devices for the same variable.
Birth weight is an especially important prognostic variable. We finely
balanced the indicator of birth weight $<2500$ grams in Table
\ref{tabfine}
using a constraint of the form (\ref{eqfinebalance}). We limited the
difference in means of birth weight in Table
\ref{tabcovs}
using a pair of constraints of the form (\ref{eqmeanmatch2}), and we limited
the number of times individual pairs $ ( \ell,m ) $ were
mismatched for the indicator of birth weight $<2500$ grams using a
constraint of the form (\ref{eqNearExact}).

\subsubsection{\texorpdfstring{Optimal selection of a
subset.}{Optimal selection of a subset}}\label{sssOptimalSubset}

Recall that our match discards some babies and must optimally decide
the following: (i)
how many babies to discard, (ii)~which babies to discard, and (iii) how to
pair the babies not discarded. Extending the technique in \citet{Ros12}
to nonbipartite matching, the objective function $\bolds{\eta}^{T}%
\mathbf{a}$ is
%
%
\begin{equation}
\label{eqObjective} \sum_{\ell=1}^{L-1}\sum
_{m=\ell+1}^{L}a_{\ell m}
\omega_{\ell
m}-\lambda\sum_{\ell=1}^{L-1}
\sum_{m=\ell+1}^{L}a_{\ell m}
\end{equation}
or $\eta_{\ell m}=\omega_{\ell m}-\lambda$, where $\omega_{\ell m}$
is a robust Mahalanobis distance between the covariates for babies
$\ell$ and $m$, and $\lambda$ is a constant selected by the
investigator. For discussion of the use of Mahalanobis distances in
matching, see \citet{Rub79}, and for a robust Mahalanobis
distance, see
\citet{Ros10}, Sections 8.3 and 13.11. Because $\sum_{\ell
=1}^{L-1}\sum_{m=\ell+1}^{L}a_{\ell m}$ is the number of matched
pairs, the objective function (\ref{eqObjective}) has the following
interpretation. When comparing two possible matched samples, say,
$a_{\ell m}$ and $a_{\ell m}^{{\prime}}$, that satisfy the constraints
with the same number of pairs, (\ref{eqObjective}) prefers the
pairing with the smaller total distance within pairs. Suppose, instead,
$%
a_{\ell m}$ includes $A>0$ more pairs than $a_{\ell m}^{{\prime}}$, $%
A=\sum_{\ell=1}^{L-1}\sum_{m=\ell+1}^{L}a_{\ell m}-a_{\ell
m}^{{\prime}}$. Then (\ref{eqObjective}) prefers $a_{\ell
m}^{{\prime}}$ to $a_{\ell m} $ if
\[
\sum_{\ell=1}^{L-1}\sum
_{m=\ell+1}^{L}a_{\ell m}^{{\prime}}
\omega_{\ell m}-\lambda\sum_{\ell=1}^{L-1}
\sum_{m=\ell+1}^{L}a_{\ell
m}^{{\prime}}<
\sum_{\ell=1}^{L-1}\sum
_{m=\ell+1}^{L}a_{\ell m} \omega_{\ell m}-
\lambda\sum_{\ell=1}^{L-1}\sum
_{m=\ell+1}^{L}a_{\ell m}
\]
or, equivalently, if
%
%
\begin{equation}
\label{eqLambdaExplained} \frac{\sum_{\ell=1}^{L-1}\sum_{m=\ell
+1}^{L}a_{\ell m} \omega_{\ell
m}-\sum_{\ell=1}^{L-1}\sum_{m=\ell+1}^{L}a_{\ell m}^{{\prime}} \omega
_{\ell m}}{\sum_{\ell=1}^{L-1}\sum_{m=\ell+1}^{L}a_{\ell m}-\sum_{\ell
=1}^{L-1}\sum_{m=\ell+1}^{L}a_{\ell m}^{{\prime}}}>\lambda.
\end{equation}
In words, the match represented by $a_{\ell m}$ had $A$ pairs more than
the match $a_{\ell m}^{{\prime}}$, so the sum of the distances $\omega
_{\ell m}$ for $a_{\ell m}$ contained $A$ more distances, and the total
distance within pairs rose by more than $A\lambda$ if (\ref
{eqLambdaExplained}) holds, so the average cost of these $A$ additional
pairs was more than $\lambda$. The objective (\ref{eqObjective}) prefers
more pairs to fewer pairs if, on average, more pairs may be had for less
than $\lambda$ and prefers fewer pairs if, on average, they cost more
than $%
\lambda$. Because $a_{\ell m}$ and $a_{\ell m}^{{\prime}}$ pair babies
differently, the change in average cost is produced by all of the paired
babies, not just $A$ babies; see \citet{Ros12} for detailed discussion.
In our case, $\lambda$ was the median of all distances before
matching, and
the algorithm prefers more pairs to fewer pairs providing the added pairs
are, on average, closer than pairs typically are. Of 231,831 babies, this
value of $\lambda$ paired 161,200 babies. Although it would be
possible to
pair additional babies, each of these additions would, on average,
raise the
distance by more than $\lambda$, that is, by more than the median pairwise
distance before matching. One might choose a different $\lambda$ in a
different context.

\subsection{\texorpdfstring{Comparison with three other
matched samples.}{Comparison with three other matched samples}}

Table
\ref{tabCompare}
compares the match described in Section~\ref{secMatchedComparison} with three
other sets of matched pairs. As noted in Section \ref
{ssMatchingTechnique}, the
match in Section~\ref{secMatchedComparison} insisted on a separation of
12 hours
in anticipated length-of-stay within each pair. Table
\ref{tabCompare}
contrasts matching with 12 hour separation to matching with no required
separation, $\geq$9 hours and $\geq$15 hours. Two quantities are
reported in Table
\ref{tabCompare}: the number of pairs and the percent of babies staying
more than one day,
where one day is a length-of-stay between 12 and 36 hours. With $0$
separation, there is only a 6.2\% difference between long-HOB and short-HOB
births in stays more than one day. With 12 hours of separation, the
difference is more than twice as large, 13.4\%. In the terminology of
\citet{AngImbRub96}, the percent of compliers is estimated to
be more than twice as large with 12 hours of separation as with 0 separation.

%
\begin{table}
\tablewidth=260pt
\caption{Comparison of four the actual match required samples with different required
differences in anticipated length of stay.
The actual matched required a 12 hour difference in anticipated length
of stay. This 12 hour required difference
is compared with 0, 9 and 15 hours.
The table records the percent of babies staying longer than
one day and the number of pairs. Because zero days is a length of stay
less than 12 hours, and one day is a length of stay
greater than 12 hours but less than 36 hours, the table indicates the
percent of babies staying longer~than 36 hours}\label{tabCompare}
\begin{tabular*}{\tablewidth}{@{\extracolsep{\fill}}ld{2.1}d{2.1}d{2.1}d{2.1}@{}}
\hline
& \multicolumn{4}{c@{}}{\textbf{Separation in anticipated LOS}} \\
\hline
Hours & 0 & 9 & 12 & 15 \\
[3pt]
Long-HOB \% & 46.9 & 50.2 & 52.7 & 58.7 \\
Short-HOB \% & 40.7 & 38.0 & 39.2 & 41.8 \\
[3pt]
Difference \% & 6.2 & 12.1 & 13.4 & 16.9 \\
[3pt]
Number of pairs & \multicolumn{1}{c}{91,053}
& \multicolumn{1}{c}{90,360} & \multicolumn{1}{c}{80,600}
& \multicolumn{1}{c@{}}{59,678} \\
\hline
\end{tabular*}
\end{table}

Matching is part of the design of an observational study, a task that should
be completed before outcomes are examined [\citet{LanRub08},
\citet{Ros10}], and, in particular, one matched sample should be
selected as
the design without using or examining outcomes. We selected the 12 hour
match based on its qualities as a matched comparison, for instance, the
covariate balance in Tables~\ref{tabl1}--\ref{tabDays} and Figure \ref
{figur1}, and the number of pairs and
instrument strength in Table
\ref{tabCompare}. The analysis of outcomes for this selected match is
discussed in Section~\ref{secInference}.

\section{\texorpdfstring{Inference: Effects on rapid
readmission.}{Inference: Effects on rapid readmission}}\label{secInference}

\subsection{\texorpdfstring{Null hypotheses of no effect or
substantial inequivalence.}{Null hypotheses of no effect or substantial
inequivalence}}

We will conduct both a test of no effect and an equivalence test for
readmissions within two days of discharge from the hospital. That is, we
wish to ask whether our data are compatible with no effect or substantial
effects of shifting the norm for length of stay. Following \citet
{BauKie96}, a three part null hypothesis is tested, where one part
asserts no effect, a second part asserts moderately large benefits from a
2-day norm and the third part asserts moderately large benefits from a
\mbox{1-day} norm. Because these three null hypotheses are logically incompatible\vadjust{\goodbreak}
with one another, at most one of the null hypotheses is true, so all three
hypotheses may be tested without a correction for testing multiple
hypotheses; see \citet{BauKie96}. In particular, the hypothesis of
no effect is a two-sided hypothesis saying changing the hour of birth
for a
baby would not change whether the baby is readmitted within two days of
discharge. The hypothesis that a norm of a one-day length-of-stay is
harmful asserts that it caused at least 500 readmissions that would not have
occurred with a two-day norm. Because there are $80\mbox{,}600$ pairs in Table
\ref{tabReadmit}, each pair containing one short-HOB baby, 500
readmissions is slightly more
%
%
\begin{table}
\tablewidth=220pt
\caption{Readmission within two days of discharge in matched pair. The
table counts pairs, not babies}\label{tabReadmit}
\begin{tabular*}{\tablewidth}{@{\extracolsep{\fill}}lcc@{}}
\hline
& \multicolumn{2}{c@{}}{\textbf{Observed data}} \\ [-4pt]
& \multicolumn{2}{c@{}}{\hrulefill}\\
& \multicolumn{2}{c@{}}{\textbf{Short-HOB baby}} \\[-4pt]
& \multicolumn{2}{c@{}}{\hrulefill}\\
\textbf{Long-HOB baby} & \multicolumn{1}{c}{\textbf{Not readmitted}} &
\textbf{Readmitted} \\
\hline
Not readmitted & 78\mbox{,}431 & 1032 \\
Readmitted & \hphantom{,0}1108 & \hphantom{00}29 \\
\hline
\end{tabular*}
\end{table}
than one half of one percent of these babies (actually $500/80\mbox{,}600=0.00620$).
In Table~\ref{tabDays}, $18\mbox{,}746-8704=10\mbox{,}042$ more long-HOB babies
stayed 2 days
rather than 1 day,
and 500 babies is about 5\% of these 10,042 babies (actually $%
500/10\mbox{,}042=0.0498$). The same value, 500, is used to test the third
hypothesis of substantial harm, rather than substantial benefit, from a
two-day norm. In testing these hypotheses, we are concerned about both
sampling variability and bias from nonrandom treatment assignment.

\subsection{\texorpdfstring{Randomization inference in matched pairs:
Viewing hour of birth
as random.}{Randomization inference in matched pairs: Viewing hour of birth
as random}}\label{ssRandInf}

There are $I$ matched pairs, $i=1,\ldots,I$ of two babies, $j=1,2$, one
treated, $Z_{ij}=1$, the other control, $Z_{ij}=0$, so
$Z_{i1}+Z_{i2}=1$ for
each $i$. In Section~\ref{ssIntroExample}, there were $I=80\mbox{,}600$ pairs of
babies, or $2\times80\mbox{,}600=161\mbox{,}200$ babies in total, and
somewhat arbitrarily
we designate short-HOB as treatment and long-HOB as control. Babies were
matched for an observed covariate $\mathbf{x}_{ij}$, so $\mathbf
{x}_{i1}=%
\mathbf{x}_{i2}$ for all $i$, but they may have differed in terms of an
unmeasured covariate $u_{ij}$, so quite possibly $u_{i1}\neq u_{i2}$ for
many or all $i$. Write $\mathbf{Z}= ( Z_{11},\ldots,Z_{I2} )^{T} $ for
the $2I$-dimensional vector of treatment assignments and
write $%
\mathcal{Z}$ for the set containing the $2^{I}$ possible values $\mathbf{z}$
of $\mathbf{Z}$, so $\mathbf{z}\in\mathcal{Z}$ if $\mathbf{z}= (
z_{11},\ldots,z_{I2} )^{T}$ with $z_{ij}=0$ or $z_{ij}=1$ and $%
z_{i1}+z_{i2}=1$ for each $i$. If $\mathcal{S}$ is a finite set, write $
\llvert\mathcal{S}\rrvert$ for the number of elements of
$\mathcal{S%
}$, so $\llvert\mathcal{Z}\rrvert=2^{I}$. Conditioning on the event
$\mathbf{Z}\in\mathcal{Z}$ is abbreviated to conditioning
on~$\mathcal{Z}$.

Each baby has two potential binary 1 or 0 responses, $r_{Tij}$ if
treated, $%
r_{Cij}$ if control, so the effect of the treatment on this baby,\vadjust{\goodbreak}
namely, $%
\delta_{ij}=r_{Tij}-r_{Cij}$, is not seen for any baby $ij$ but the
response actually seen from $ij$ is $R_{ij}=Z_{ij} r_{Tij}+ (
1-Z_{ij} ) r_{Cij}=r_{Cij}+Z_{ij} \delta_{ij}$; see \citet{Spe90},
\citet{Wel37}, \citet{Rub74}, \citet{Rei00} or
\citet{Gad01}.
Write $%
\mathbf{R}= ( R_{11},\ldots,R_{I2} )^{T}$, $\bolds{\delta
}%
= ( \delta_{11},\ldots,\delta_{I2} )^{T}$, $\mathbf{r}%
_{C}= ( r_{C11},\ldots,r_{CI2} )^{T}$, $\mathbf{r}_{T}= (
r_{T11},\ldots,r_{TI2} )^{T}$, so $\mathbf{r}_{T}=\mathbf{r}_{C}+%
\bolds{\delta}$. Here, $\delta_{ij}\in\{ -1,0,1 \} $ for
each $ij$ and Fisher's (\citeyear{Fi35}) sharp null hypothesis $H_{0}$
of no treatment
asserts that $H_{\mathbf{0}}:\bolds{\delta}=\mathbf{0}$. In the
discussion here, $R_{ij}$ indicates whether baby $ij$ was readmitted, $%
R_{ij}=1$, or not, $R_{ij}=0$, within two days of discharge from the
hospital. If $r_{Tij}=1$ and $r_{Cij}=0$ so $\delta
_{ij}=r_{Tij}-r_{Cij}=1 $, then baby $ij$ would have been readmitted if born
at an hour that would typically lead to a one-day stay and would not have
been readmitted if born at an hour that would typically lead to a two-day
stay, so being born at a short-HOB rather than a long-HOB would have caused
this baby to be readmitted. Aside from Fisher's null hypothesis of no
effect, greatest interest attaches to hypotheses in which one treatment may
cause but does not prevent a readmission, $H_{\bolds{\delta}_{0}}:%
\bolds{\delta}=\bolds{\delta}_{0}$ with $\bolds{\delta}%
_{0}\geq\mathbf{0}$ and $\bolds{\delta}_{0}\neq\mathbf{0}$, because
hypotheses of this form say that one treatment is clearly better than the
other. Write $\mathcal{F}= \{ ( r_{Tij},r_{Cij},\mathbf{x}%
_{ij},u_{ij} ), i=1,\ldots,I, j=1,2 \} $ for the potential
responses and covariates.

In a paired randomized experiment, one baby in each pair would be
picked at
random for treatment, the other baby receiving control, with independent
assignments in distinct pairs, that is, $\Pr( \mathbf{Z}=\mathbf
{z}%
\vert\mathcal{F}, \mathcal{Z} ) =2^{-I}$ for $\mathbf{z%
}\in\mathcal{Z}$. In Section~\ref{ssIntroExample}, hour-of-birth is not
randomized, but because hour of birth should not pick out a particular type
of baby, the hope is that $\Pr( \mathbf{Z}=\mathbf{z}\vert
\mathcal{F}, \mathcal{Z} ) $ is close to the randomization
distribution. Section~\ref{ssSensitivity} examines the sensitivity of
conclusions to departures of various magnitudes from $\Pr( \mathbf
{Z}=%
\mathbf{z} \vert\mathcal{F}, \mathcal{Z} ) =2^{-I}$.

The statistic $T=\sum_{i=1}^{I}\sum_{j=1}^{2}Z_{ij} R_{ij}$ is the observed
number of readmissions within two days among babies born at a short-HOB.
Some of the readmissions recorded in $T$ may have been caused by the
short-HOB and others might have occurred whether the baby was born at a
short or a long HOB. The unobservable quantity $T_{c}=\sum_{i=1}^{I}%
\sum_{j=1}^{2}Z_{ij} r_{Cij}$ is the number of readmissions that would have
occurred had all babies been born at a long-HOB. Fisher's sharp null
hypothesis, $H_{\mathbf{0}}:\bolds{\delta}=\mathbf{0}$, says that no
readmission was caused or prevented by the hour of birth, with the
consequence that $T=T_{c}$. Consider the distribution of $T_{c}$ in a
randomized experiment, that is, $\Pr( T_{c}\leq k \vert
\mathcal{F}, \mathcal{Z} ) $ when $\Pr( \mathbf{Z}=%
\mathbf{z} \vert\mathcal{F}, \mathcal{Z} ) =2^{-I}$.
Define $n_{11}$ to be the number of pairs $i$ with
$r_{Ci1}=r_{Ci2}=1$, $%
n_{00}$ to be the number of pairs with $r_{Ci1}=r_{Ci2}=0$, and
$n_{10}$ to
be the number of pairs with $r_{Ci1}\neq r_{Ci2}$. If $H_{\mathbf{0}}:%
\bolds{\delta}=\mathbf{0}$ were true, then $R_{ij}=r_{Cij}$ and it
would be possible to calculate $ ( n_{11},n_{10},n_{00} ) $ from
the observed $R_{ij}$'s. Because $\Pr( \mathbf{Z}=\mathbf{z}%
\vert\mathcal{F}, \mathcal{Z} ) =2^{-I}$ and $\mathbf{r%
}_{C}$ is fixed by conditioning on $\mathcal{F}$, the $I$ terms $%
\sum_{j=1}^{2}Z_{ij} r_{Cij}$ are\vspace*{1pt} independent for distinct $i$, and $%
\sum_{j=1}^{2}Z_{ij} r_{Cij}$ is 1 with certainty if the pair is concordant
with $r_{Ci1}=r_{Ci2}=1$, is 0 with certainty if the pair is concordant with
$r_{Ci1}=r_{Ci2}=0$, and is 1 or 0 each with probability $\frac{1}{2}$ if
the pair is discordant with $r_{Ci1}\neq r_{Ci2}$; therefore, $T_{c}$
is the
constant $n_{11}$ plus a binomial random variable with probably of
success $%
\frac{1}{2}$ and sample size $n_{10}$. Because $T=T_{c}$ when Fisher's
sharp null hypothesis $H_{\mathbf{0}}:\bolds{\delta}=\mathbf{0}$ is
true, it follows that $H_{0}$ may be tested in a randomized experiment by
comparing $T$ with the randomization distribution of $T_{c}$, and this is
essentially the same as McNemar's test.

Let $\bolds{\delta}_{0}$ be a $2I$-dimensional with coordinates $%
\delta_{0ij}\in\{ -1,0,1 \} $, and consider the hypothesis
$H_{%
\bolds{\delta}_{0}}:\bolds{\delta}=\bolds{\delta}_{0}$.
Not all hypotheses of this form are logically compatible with the observed
data because $R_{ij}-Z_{ij} \delta_{ij}=r_{Cij}$ and $R_{ij}+ (
1-Z_{ij} ) \delta_{ij}=r_{Tij}$ must both be in $ \{ 0,1
\}
$. If $H_{\bolds{\delta}_{0}}$ is logically incompatible with the
data, we may reject it with type 1 error rate of zero, so for the remainder
of the discussion, assume that $H_{\bolds{\delta}_{0}}$ is logically
compatible with the observed data, or briefly compatible. If $H_{%
\bolds{\delta}_{0}}:\bolds{\delta}=\bolds{\delta}_{0}$
were true (and hence compatible), then $r_{Cij}=R_{ij}-Z_{ij} \delta_{0ij}$
may be calculated from the hypothesis and the data, so $n_{11}$,
$n_{10}$, $%
n_{00}$ and $T_{c}$ may be calculated as well, so $T_{c}$ may be compared
with the constant-plus-binomial distribution to test $H_{\bolds
{\delta}%
_{0}}$. Unfortunately, there are many hypotheses $H_{\bolds{\delta}
_{0}}:\bolds{\delta}=\bolds{\delta}_{0}$ and it is not practical
to test them all; however, the testing of many hypotheses
$H_{\bolds{%
\delta}_{0}}:\bolds{\delta}=\bolds{\delta}_{0}$ may be
summarized using a scalar quantity, the attributable effect.

The attributable\vspace*{1pt} effect $\Delta=\sum_{i=1}^{I}\sum_{j=1}^{2}Z_{ij}
\delta_{ij}$ is an unobservable quantity giving the net increase in the
number of
babies readmitted because they were born at a short-HOB; see \citet
{Ros95}. It is a random variable because it depends upon $\mathbf{Z}$, but
it is not an observable random variable because it depends on
$\bolds{%
\delta}$. Among\vspace*{-1pt} babies born at a short-HOB, we see $T=\sum_{i=1}^{I}%
\sum_{j=1}^{2}Z_{ij} R_{ij}=\sum_{i=1}^{I}\sum_{j=1}^{2}Z_{ij} r_{Tij}$
readmissions,\vspace*{1pt} whereas these same babies would have had $T_{c}=\sum
_{i=1}^{I}%
\sum_{j=1}^{2}Z_{ij} r_{Cij}$ readmissions had they been born at a
long-HOB. If $H_{\bolds{\delta}_{0}}:\bolds{\delta}=%
\bolds{\delta}_{0}$ were true, then $\Delta$ may be calculated using
the hypothesized $\bolds{\delta}_{0}$ as $\Delta_{0}=\sum
_{i=1}^{I}\sum_{j=1}^{2}Z_{ij} \delta_{0ij}$, and $T-\Delta_{0}$
would equal $T_{c}$.

For the reason noted above, we consider hypotheses $H_{\bolds
{\delta}%
_{0}}:\bolds{\delta}=\bolds{\delta}_{0}$ that say that one
treatment is better than the other in the sense that $\bolds{\delta
}%
_{0}\geq\mathbf{0}$ and $\bolds{\delta}_{0}\neq\mathbf{0}$. We will
do this twice, once reversing the roles of treatment and control, but for
the moment consider the hypothesis that a short-HOB may cause but not
prevent readmissions in the sense that $\bolds{\delta}_{0}\geq
\mathbf{%
0}$. A value of $\Delta_{0}$ is rejected if every hypothesis $H_{%
\bolds{\delta}_{0}}:\bolds{\delta}=\bolds{\delta}_{0}$
with $\bolds{\delta}_{0}\geq\mathbf{0}$ and $\bolds{\delta}%
_{0}\neq\mathbf{0}$ that gives rise to this value of $\Delta_{0}=\sum
_{i=1}^{I}\sum_{j=1}^{2}Z_{ij} \delta_{0ij}$ is rejected;
otherwise, this value of $\Delta_{0}$ is not rejected. For all of
these hypotheses, $T-\Delta_{0}=T_{c}$ will be the same number;
however, $n_{11}$, $n_{10}$ and $n_{00}$ typically change with
$\bolds{\delta}_{0}$.
For a given $\Delta_{0}$, among all hypotheses $H_{\bolds{\delta}%
_{0}}:\bolds{\delta}=\bolds{\delta}_{0}$ with $\bolds{%
\delta}_{0}\geq\mathbf{0}$ and $\bolds{\delta}_{0}\neq\mathbf{0}$
that yield the same attributable effect $\Delta_{0}$, there is one
hypothesis $H_{\widetilde{\bolds{\delta}}_{0}}:\bolds{\delta}=
\widetilde{\bolds{\delta}}_{0}$ with $\Delta_{0}=\sum_{i=1}^{I}\sum
_{j=1}^{2}Z_{ij} \widetilde{\delta}_{0ij}$ that is
the most difficult\vspace*{1pt} to reject, so if $H_{\widetilde{\bolds{\delta}}_{0}}
$ is rejected, then the associated value of $\Delta_{0}$ is rejected.
In a\vspace*{1pt}
cohort study, as in Section~\ref{ssIntroExample}, this hypothesis
$H_{\widetilde{%
\bolds{\delta}}_{0}}:\bolds{\delta}=\widetilde{\bolds{%
\delta}}_{0}$ has $\sum_{j=1}^{2}Z_{ij} \delta_{ij}=1$ for\vspace*{1pt} as many
pairs with \mbox{$R_{i1}+R_{i2}=2$} as possible; see Rosenbaum
[(\citeyear{Ros95}), Section~6] for a precise statement and proof. For
instance, if Table~\ref{tabReadmit} had come from a randomized
experiment, $\Pr( \mathbf{Z}=\mathbf{z} 
\vert\mathcal{F}, \mathcal{Z} ) =2^{-I}$, then $\Delta_{0}\geq500$
would be rejected if McNemar's one-sided test rejected no
effect in the adjusted Table~\ref{tabReadmitAdj}, where all 29 pairs with $R_{i1}+R_{i1}=2$ have
$\widetilde{\delta}_{0ij}=1
$ and 471 pairs with $R_{i1}+R_{i1}=1$ have $\widetilde{\delta}_{0ij}=1$.
Why is this $H_{\widetilde{\bolds{\delta}}_{0}}$ the hypothesis that
is most difficult to reject among hypotheses with $\Delta_{0}\geq500$?
Intuitively, this $H_{\widetilde{\bolds{\delta}}_{0}}$ has $\Delta
_{0}=500$ with the most variability because the number of discordant
pairs $%
n_{10}$ is as large as possible; see Rosenbaum [(\citeyear{Ros95}),
Section 6] for precise discussion.

%
\begin{table}
\tablewidth=240pt
\caption{Readmission within two days of discharge in matched pair
adjusted for a null hypothesis $H_{0}\dvtx\bolds{\delta} = \bolds
{\delta}_{0}$ that
attributes $\sum\delta_{ij} Z_{ij} = \Delta=500$ readmissions to early
discharge}\label{tabReadmitAdj}
\begin{tabular*}{\tablewidth}{@{\extracolsep{\fill}}lcc@{}}
\hline
& \multicolumn{2}{c@{}}{\textbf{Data adjusted for} $\bolds{H_{0}\dvtx\bolds{\delta} =
\bolds{\delta}_{0}}$} \\[-4pt]
& \multicolumn{2}{c@{}}{\hrulefill}\\
& \multicolumn{2}{c@{}}{\textbf{Short-HOB baby}} \\[-4pt]
& \multicolumn{2}{c@{}}{\hrulefill}\\
\textbf{Long-HOB baby} & \textbf{Not readmitted}
& \textbf{Readmitted} \\
\hline
Not readmitted& 78,902 & 561 \\
Readmitted & \hphantom{0}1137 & \hphantom{00}0 \\
\hline
\end{tabular*}
\end{table}

If Table
\ref{tabReadmit}
had been seen in a randomized experiment, $\Pr( \mathbf{Z}=\mathbf
{z}%
\vert\mathcal{F}, \mathcal{Z} ) =2^{-I}$, then the procedure just
described would yield the following conclusions. Testing
the null hypothesis of no effect, $H_{\mathbf{0}}:\bolds{\delta}=%
\mathbf{0}$, yields a two-sided $P$-value of 0.105 using McNemar's two-sided
test, so no effect is plausible. Is a substantial benefit of $\Delta
_{0}=500$ from being born at a long-HOB also plausible? It is not.
McNemar's one-sided test rejects in Table
\ref{tabReadmitAdj}
with $P$-value $2.1\times10^{-45}$, so it rejects for every
$H_{\bolds{%
\delta}_{0}}:\bolds{\delta}=\bolds{\delta}_{0}$ with $%
\bolds{\delta}_{0}\geq\mathbf{0}$ and $\bolds{\delta}_{0}\neq
\mathbf{0}$ and $\Delta_{0}\geq500$. Reversing the roles of (and
notation for) a short-HOB and a long-HOB, a substantial benefit of
$\Delta_{0}=500$ from being born at a short-HOB is rejected with a
$P$-value $%
2.9\times10^{-25}$. In brief, if Table
\ref{tabReadmit}
had been seen in a randomized experiment, the hypothesis of no effect would
be plausible, whereas a benefit or harm that affected at least one half of
one percent of babies would not be remotely plausible. Of course, Table
\ref{tabReadmit}
is not from a randomized experiment.

Our hope has been that a baby's hour of birth tells you little or nothing
about the baby and her mother, that is, our hope was that hour of birth was
nearly random, at least after matching for covariates. We cannot be
certain of this, however. It is possible to use drugs to induce or
accelerate labor, and perhaps the use of such drugs shifts the hour of
delivery for some mothers, possibly in a fashion that biases randomization
inferences based on Table
\ref{tabReadmit}. Moreover, the distribution of times for vaginal
delivery may be affected
by cesarean sections, which again may be related to aspects of the
mother or
the hospital. How large would such biases have to be to alter the
qualitative conclusions based on randomization inferences? This is
examined in Section~\ref{ssSensitivity} using a sensitivity
analysis.\

\subsection{\texorpdfstring{Sensitivity analysis in matched pairs: What
if birth hour
is not random\textup{?}}{Sensitivity analysis in matched pairs: What if
birth hour
is not random}}\label{ssSensitivity}

The assumption in Section~\ref{ssRandInf} was that hour of birth is effectively
random, that it tells you nothing about the baby or the mother or the
hospital and its staff, so that $\Pr( \mathbf{Z}=\mathbf{z}%
\vert\mathcal{F}, \mathcal{Z} ) =2^{-I}$ for $\mathbf{z%
}\in\mathcal{Z}$. The current section studies sensitivity of the
conclusions to quantified violations of this assumption. The model (\ref
{eqsenmod}) for sensitivity analysis used here is discussed in
\citet{Ros02}, Section 4. Other methods of sensitivity analysis in
observational
studies are discussed by \citet{CORetal59}, \citet{RosRub},
\citet{Yan84}, \citet{Gas92}, \citet{Mar}, \citet{Imb03},
\citet{DipGan04}, \citet{YuGas05}, \citet{WanKri06},
\citet{McCGusLev07}, \citet{EglSchMac09} and \citet{HosHanHol10},
among others.

One model for sensitivity analysis in observational studies asserts
that, in
the population before matching, treatment assignments are independent and
two babies, say, $ij$ and $ij^{\prime}$, with the same observed
covariates, $%
\mathbf{x}_{ij}=\mathbf{x}_{ij^{\prime}}$, may differ in their odds of
treatment by at most a factor of $\Gamma\geq1$,
%
%
\begin{equation}
\label{eqsenmodOR} \frac{1}{\Gamma}\leq\frac{\Pr( Z_{ij}=1 \vert%
\mathcal{F} ) \Pr( Z_{ij^{\prime}}=0
\vert\mathcal{F} )}{\Pr( Z_{ij}=0
\vert\mathcal{F} ) \Pr( Z_{ij^{\prime}}=1
\vert\mathcal{F} )}\leq\Gamma;
\end{equation}
then the distribution of $\mathbf{Z}$ is returned to $\mathcal{Z}$ by
conditioning on $\mathbf{Z}\in\mathcal{Z}$. Model (\ref{eqsenmodOR}) is
similar to the sensitivity analysis of \citet{CORetal59} and is
exactly the same as assuming that
%
%
\begin{eqnarray}
\label{eqsenmod} \Pr( \mathbf{Z}=\mathbf{z}\vert\mathcal{F},
\mathcal{Z}%
) &=&\prod_{i=1}^{I}
\frac{\exp( \sum_{j=1}^{2}\gamma
z_{ij} u_{ij} )}{\exp( \gamma u_{i1} ) +\exp(
\gamma u_{i2} )}\hspace*{-30pt}
\\
&=&\frac{\exp( \gamma\mathbf{u}^{T}\mathbf{z} )}{\sum_{\mathbf{b%
}\in\mathcal{Z}}\exp( \gamma\mathbf{u}^{T}\mathbf{b} )}
\qquad\mbox{with }\mathbf{u}\in[ 0,1 ]^{2I}\mbox{ and }
\gamma=\log( \Gamma);\hspace*{-30pt}
\end{eqnarray}
see Rosenbaum [(\citeyear{Ros02}), Section 4] where $u_{ij}$ satisfying
(\ref
{eqsenmod}) is
constructed from $\Pr( Z_{ij}=1 \vert\mathcal{F}%
) $ satisfying (\ref{eqsenmodOR}) and conversely.

Using either of the two approaches in \citet{GasKriRos98}
or \citet{RosSil09N2}, the one parameter $\Gamma$ may be unpacked
into two
sensitivity parameters, one controlling the relationship between $u_{ij}$
and treatment $Z_{ij}$, the other controlling the relationship between $
u_{ij}$ and response under control $r_{Cij}$. For instance, an unobserved
covariate $u_{ij}$ that both doubles the odds of a short-HOB and
doubles the
odds of readmission within two days is equivalent to $\Gamma=1.25$, whereas
doubling the odds of a short-HOB with a four-fold increase in the odds of
readmission is equivalent to $\Gamma=1.5$. See \citet{GasKriRos98}
and \citet{RosSil09N2} for specifics, noting that the approaches
taken in these two papers differ in general but agree in the case of a
binary outcome $r_{Cij}$. See \citet{Gas92} for related results
for the
method of \citet{CORetal59}.

Under (\ref{eqsenmodOR}) or (\ref{eqsenmod}), sharp lower and upper bounds
on the distribution of $T_{c}$ are obtained as a constant plus a binomial
random variable with $n_{10}$ trials and, respectively, probabilities $%
1/ ( 1+\Gamma) $ and $\Gamma/ ( 1+\Gamma) $, yielding
an interval of possible $P$-values for each $\Gamma\geq1$; see
\citet{Ros95}. Consider the null hypothesis that being born at a short-HOB
sometimes causes but never prevents readmission within two days such
that at
least $\Delta_{0}=500$ readmissions were caused. Testing the null
hypothesis $\Delta_{0}\geq500$, the upper bound on the $P$-value is 0.040
for $\Gamma=1.85$ and 0.110 for $\Gamma=1.9$. Reversing roles and
testing the less plausible null hypothesis that a long-HOB causes but does
not prevent readmissions and caused at least 500 readmissions, the upper
bound on the $P$-value is 0.0192 for $\Gamma=1.5$ and 0.079 for $\Gamma
=1.55$. In brief, for it to be plausible that $\Delta_{0}=500$
readmissions were caused or prevented by short-versus-long-HOB, the
unobserved covariate $u_{ij}$ would need a $\Gamma>1.5$. As mentioned in
the previous paragraph, a $\Gamma=1.5$ corresponds with a $u_{ij}$ that
doubles the odds of delivering at a long-HOB and increases the odds of
readmission by a factor of four.

\section{\texorpdfstring{Summary: Flexible new tools for nonbipartite
matching.}{Summary: Flexible new tools for nonbipartite matching}}

When compared with network optimization [e.g., \citet{Der88}], the integer
programming formulation in Section~\ref{secInteger} substantially
enlarges the
set of tools available for nonbipartite matching to strengthen an
instrumental variable. Among the new tools not previously available are
the following:
(i) fine or near-fine nonbipartite matching for one or more nominal
variables (\ref{eqfinebalance}), (ii) nonbipartite matching with constraints
on imbalances in means (\ref{eqmeanmatch1}), and (iii) optimal subset
nonbipartite matching using~(\ref{eqObjective}), (iv) combining fine balance
with optimal subset nonbipartite matching. In the example, this approach
formed 80,600 pairs of two babies who were similar on numerous covariates
yet very different in anticipated length of stay based on hour of birth.


%

\printaddresses

\end{document}